\documentclass[a4paper,twocolumn,accepted=2020-09-08]{quantumarticle}
\pdfoutput=1

\usepackage{amsmath,amssymb,amsfonts,amsthm}
\usepackage{mathtools}
\usepackage{bbm}
\usepackage{bm}

\usepackage[utf8]{inputenc}
\usepackage[T1]{fontenc}
\usepackage{accents}
\usepackage[colorlinks]{hyperref}
\usepackage[normalem]{ulem}

\usepackage{graphicx}

\usepackage[numbers,sort&compress]{natbib}

\def\bra#1{\mathinner{\langle{#1}|}}
\def\ket#1{\mathinner{|{#1}\rangle}}
\def\braket#1{\mathinner{\langle{#1|#1}\rangle}}
\def\ketbra#1{\mathinner{|{#1}\rangle\!\langle{#1}|}}

{\catcode`\|=\active 
  \gdef\Braket#1{\left<\mathcode`\|"8000\let|\BraVert {#1}\right>}}
\def\BraVert{\egroup\,\mid@vertical\,\bgroup}
\newcommand{\oprod}[2]{| #1 \rangle\!\langle #2 |}
\newcommand{\iprod}[2]{\langle #1 | #2 \rangle}

\def\dbra#1{\mathinner{\langle\!\langle{#1}|}}
\def\dket#1{\mathinner{|{#1}\rangle\!\rangle}}
\def\dbraket#1{\mathinner{\langle\!\langle{#1|#1}\rangle\!\rangle}}
\def\dketbra#1{\mathinner{|{#1}\rangle\!\rangle\!\langle\!\langle{#1}|}}

\newcommand{\diprod}[2]{\langle\!\langle #1 | #2 \rangle\!\rangle}

\DeclareMathOperator{\Tr}{Tr}
\DeclareMathOperator{\range}{range}
\renewcommand\L{\mathcal{L}}
\newcommand{\M}{\mathcal{M}}
\newcommand{\HS}{\mathcal{H}}
\newcommand{\N}{\mathcal{N}}
\newcommand{\X}{\mathcal{X}}

\newcommand{\C}{\mathcal{C}}
\newcommand{\Z}{\mathcal{Z}}
\newcommand{\Q}{\mathcal{Q}}
\newcommand{\id}{\mathbbm{1}}

\begin{document}

\title{Communication Through Coherent Control of Quantum Channels}

\author{Alastair A.\ Abbott}
\affiliation{D\'epartement de Physique Appliqu\'ee, Universit\'e de Gen\`eve, 1211 Gen\`eve, Switzerland}
\affiliation{Univ.\ Grenoble Alpes, CNRS, Grenoble INP, Institut N\'eel, 38000 Grenoble, France}
\orcid{0000-0002-2759-633X}

\author{Julian Wechs}
\affiliation{Univ.\ Grenoble Alpes, CNRS, Grenoble INP, Institut N\'eel, 38000 Grenoble, France}
\orcid{0000-0002-0395-6791}

\author{Dominic Horsman}
\affiliation{Univ.\ Grenoble Alpes, CNRS, Grenoble INP, LIG, 38000 Grenoble France}
\orcid{0000-0003-4965-0584}

\author{Mehdi Mhalla}
\affiliation{Univ.\ Grenoble Alpes, CNRS, Grenoble INP, LIG, 38000 Grenoble France}
\orcid{0000-0003-4178-5396}

\author{Cyril Branciard}
\affiliation{Univ.\ Grenoble Alpes, CNRS, Grenoble INP, Institut N\'eel, 38000 Grenoble, France}
\orcid{0000-0001-9460-825X}

\date{Submitted July 15, 2019; revised April 14, 2020; accepted September 8, 2020}

\begin{abstract}
	A completely depolarising quantum channel always outputs a fully mixed state and thus cannot transmit any information.
	In a recent Letter [D.\ Ebler \emph{et al.}, \href{https://doi.org/10.1103/PhysRevLett.120.120502}{Phys.\ Rev.\ Lett.\ \textbf{120}, 120502 (2018)}], it was however shown that if a quantum state passes through two such channels in a quantum superposition of different orders---a setup known as the ``quantum switch''---then information can nevertheless be transmitted through the channels.
	Here, we show that a similar effect can be obtained when one coherently controls between sending a target system through one of two identical depolarising channels. 
	Whereas it is tempting to attribute this effect in the quantum switch to the indefinite causal order between the channels, causal indefiniteness plays no role in this new scenario. This raises questions about its role in the corresponding effect in the quantum switch.
	We study this new scenario in detail and we see that, when quantum channels are controlled coherently, information about their specific implementation is accessible in the output state of the joint control-target system. 
	This allows two different implementations of what is usually considered to be the same channel to therefore be differentiated.
	More generally, we find that to completely describe the action of a coherently controlled quantum channel, one needs to specify not only a description of the channel (e.g., in terms of Kraus operators), but an additional ``transformation matrix'' depending on its implementation.
	\end{abstract}

\maketitle

\section{Introduction}

The ability to create superpositions of quantum states opens up many advantages for communication and information processing that are inaccessible to classical mixtures of states, exemplified by their use in controlled logic gates (e.g., \textsc{cnot}) in quantum computing~\cite{nielsen11}.
Recently, it has been shown that a coherent quantum control system can be used to even put the causal ordering of quantum channels into superposition, thus rendering it ``indefinite'', in the so-called ``quantum switch''~\cite{chiribella13}.
Surprisingly, when certain zero-capacity channels are placed in a quantum switch the resulting switched channels still allow information to be transmitted, something impossible if their causal ordering is fixed or controlled classically~\cite{ebler18}.

Motivated by this example, we revisit here the notion of coherent control of arbitrary quantum channels---something that has in the generally been considered problematic or, at best, subtle~\cite{Araujo14a,friis14,rambo16,thompson18}---by exploiting a control system to determine which channel is used to transmit a state rather than the order in which two communication channels are used.
This approach of channel multiplexing has previously been used for error filtration~\cite{gisin05} and discussed as an approach to control unknown unitaries~\cite{Araujo14a}. 
We show here that it allows one to obtain a similar, counter-intuitive communication advantage to that obtained with the quantum switch mentioned earlier, despite the absence of any ``causal indefiniteness'':
when each channel is maximally noisy, information can nonetheless be transmitted through the coherently multiplexed communication channels.

When controlled coherently in this way we find that---in contrast to the quantum switch~\cite{chiribella13,ebler18}---the action of the ``global'' multiplexed channel depends not only the descriptions of the individual channels as completely positive trace-preserving (CPTP) maps---i.e., their standard description as ``quantum channels''~\cite{Kraus83,nielsen11,wilde13}---but also on more fine-grained information about their realisations.
This includes, but goes beyond, relative phase information, highlighting the subtleties involved in describing ``controlled channels'': indeed as we will see, without extra information on the specific channel implementation the problem is in fact ill-defined.
As a result, in any framework for manipulating channels that includes their coherent control, their description must be supplemented by further information than just the corresponding CPTP map.

\section{Communication through the ``depolarising quantum switch''}

Let us first review the quantum switch and the aforementioned communication advantage obtained exploiting it.
The quantum switch is a process comprising a coherent control qubit, a $d$-dimensional target system, and a pair of ``black box'' operations that, taken individually, implement some CPTP maps---so-called ``quantum channels''---$\C_0$ and $\C_1$ on their input systems~\cite{chiribella13}.
If the control qubit is in the state $\ket{0}^c$, then first $\C_0$ then $\C_1$ is applied to the target system, while if it is in the state $\ket{1}^c$ then the operations are in the opposite order.
Initialising the control in the state $\ket{+}^c=\frac{1}{\sqrt{2}}(\ket{0}^c+\ket{1}^c)$ therefore applies the operations in a superposition of the two orders. 
Since, in this case, one cannot say that either operation is definitely applied before another, the quantum switch is said to exhibit indefinite causal order~\cite{chiribella13}.
The quantum switch is, more formally, a ``quantum supermap''~\cite{chiribella08a}, and its causal indefiniteness can be formally understood through the notion of ``causal nonseparability''~\cite{oreshkov12,araujo15,oreshkov16,wechs19}.
Several recent experiments have been conducted to implement the quantum switch and experimentally verify its causal indefiniteness~\cite{procopio15,rubino17,rubino17a,goswami18,goswami18a,wei19,guo20}.

Indefinite causal order is known to be a resource providing advantages in some tasks over any (quantum or classical) process with a definite causal order~\cite{oreshkov12,chiribella13,chiribella12,colnaghi12,araujo14,facchini15,feix15,guerin16}.
In Ref.~\cite{ebler18}, it was observed that, if the CPTP maps $\C_i$ are taken to be fully depolarising channels $\N_i$ (which map any initial target state $\rho_\text{in}^t$ to the maximally mixed state $\frac{\id^t}{d}$), then the switch (with the initial state of the control qubit fixed to $\ket{+}^c$) implements a global channel $\mathcal{S}[\N_0,\N_1]$ mapping  $\rho_\text{in}^t$ to the joint control-target state
\begin{equation}
\rho_\text{out}^{ct} = \frac{\id^c}{2} \otimes \frac{\id^t}{d} + \frac{1}{2} \big[ \oprod{0}{1}^c + \oprod{1}{0}^c \big] \otimes \frac{1}{d^2}\rho_\text{in}^t \,, \label{eq:rho_out_full_switch}
\end{equation}
which is \emph{not} $\frac{\id^t}{d}$ but instead retains some dependence on $\rho_\text{in}^t$.
Thus, information can propagate through the ``depolarising quantum switch'' despite this being impossible for the channels $\N_0$, $\N_1$, $\N_1\circ\N_0$, and $\N_0\circ\N_1$ individually.
This somewhat surprising result arising in the presence of indefinite causal order has recently been subject to experimental tests~\cite{goswami18a,guo20}, and generalised to setups that put more channels in a superposition of different orders~\cite{procopio19,procopio20}.
Note that $\mathcal{S}[\N_0,\N_1]$ above---and more generally, $\mathcal{S}[\C_0,\C_1]$---depend only on the CPTP maps implemented by the black-box operations, despite the fact the operations are applied only to a subspace of the joint target-control space~\cite{chiribella13,ebler18}; this is a consequence of the quantum switch being a quantum supermap.

\section{Communication through coherently-controlled depolarising channels}

\begin{figure}[t]
	\begin{center}
	\includegraphics[width=\columnwidth]{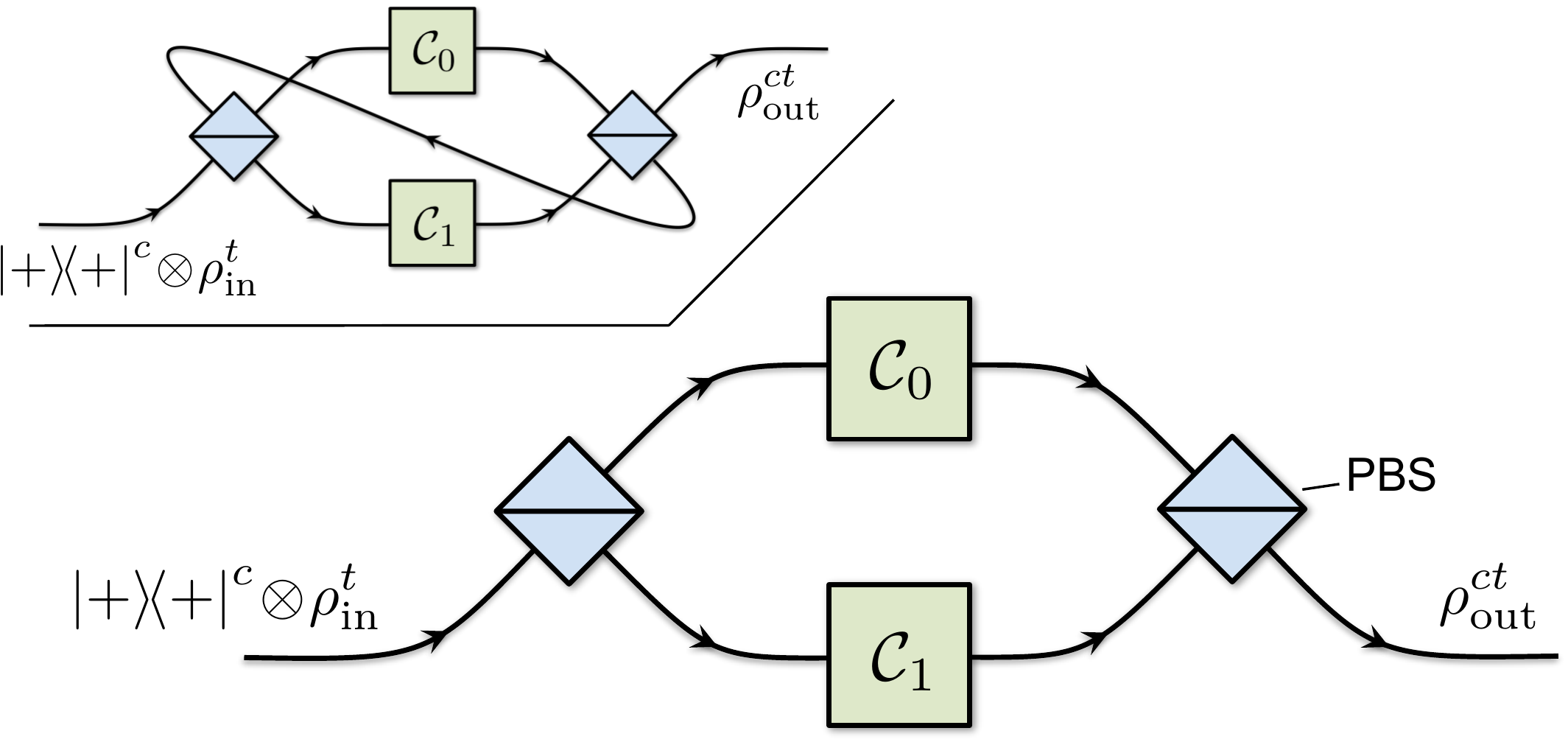}
	\end{center}
	\caption{
	The upper-left inset shows a typical photonic setup for the quantum switch~\cite{chiribella13,araujo14}, in which the control qubit is encoded in the polarisation of a photon which is routed by polarising beamsplitters (PBS), and the target system is encoded in some internal degree of freedom of the photon (as e.g.\ in Refs.~\cite{goswami18,goswami18a}). In the present work we consider only the ``first half'' of such a setup, as shown in the main figure. This implements a coherent control between the two boxes implementing $\C_0$ and $\C_1$ that the target system traverses.
	As we show, the above diagram is in fact \emph{ill-defined} since, when controlled coherently, the maps $\C_i$ do not fully determine the output state $\rho^{ct}_\text{out}$; see Fig.~\ref{fig2}.
	}
	\label{fig1}
\end{figure}

Thus far, all attempts to implement the quantum switch have relied on interferometric setups~\cite{procopio15,rubino17,rubino17a,goswami18,goswami18a,wei19,guo20,taddei20}.%
\footnote{While there has been some debate as to whether such experiments are genuine implementations of the quantum switch~\cite{oreshkov19,vojinovic19}, this issue is beyond the scope of the present paper. Here, we rather draw inspiration from such implementations to consider the case of coherently controlled channels.}
In a typical such setup, the target system is routed to the switched operations, which here correspond to communication channels, via some beamsplitters (see insert of Fig.~\ref{fig1}). 
Motivated by such an approach, in this work we consider instead the state of the joint control-target system after traversing only half of such an implementation; that is, after the target system has passed, in a superposition, through the communication channels only a single time.
This situation, a possible implementation of which is shown in Fig.~\ref{fig1}, amounts to coherently controlling between applying the operations implementing $\C_0$ or $\C_1$ to the target system. 
By preparing the control qubit in the state $\ket{+}^c$, a ``superposition'' of the two operations is thus applied.\footnote{Our calculations below generalise easily to any other initial state of the control (not necessarily $\ket{+}^c$).}

We note that in general, coherent control of completely unknown quantum operations is impossible. 
However, in interferometer-type situations the implementations of operations are localised and thus act trivially on modes that do not pass through them; 
this provides additional information about the structure of the joint control-target Hilbert space that makes such control possible~\cite{Araujo14a,friis14,rambo16,thompson18}. 
Indeed, coherent control of unitary operations by such means has been demonstrated experimentally in many scenarios~\cite{lanyon09,zhou11,zhou13,friis15,dunjko15}.  

Let us consider, as in Refs.~\cite{ebler18}, the case where the two operations implement fully depolarising channels ($\C_i=\N_i$), and consider first the concrete case analysed in Ref.~\cite{ebler18} and implemented experimentally in Refs.~\cite{goswami18a,guo20} (for the case of a qubit target system, $d=2$) where these are realised by randomising over a set of $d^2$ orthogonal unitary operators $\{U_i\}_{i=0}^{d^2-1}$. 
For each channel, one then indeed has $\N_{0/1}(\rho_\text{in}^t)=\frac{1}{d^2}\sum_iU_i\rho_\text{in}^t\, U_i^\dagger=\frac{\id^t}{d}$.

For each random choice of unitary operators $(U_i,U_j$), the control-target system therefore undergoes the unitary evolution $\ketbra{0}^c\otimes U_i+\ketbra{1}^c\otimes U_j$.\footnote{Note that $U_i$ and $U_j$ (and similarly, the Kraus operators $K_i$ and $L_j$ considered later on) must be written with respect to a common reference phase. We also assume that the arms of the circuit do not introduce any additional relative phases).}
If the control qubit is initially in the state $\ket{+}^c$ and the target system is in some state $\ket{\psi_\text{in}}^t$, the joint system thus evolves to the state
\begin{align}
\ket{\Phi_{ij}}^{ct}  = \frac{1}{\sqrt{2}}\left( \ket{0}^c\otimes U_i \ket{\psi_\text{in}}^t + \ket{1}^c\otimes U_j \ket{\psi_\text{in}}^t \right).
\end{align}
Averaging over all choices of $(U_i,U_j)$ one finds that the output state is
\begin{align}
\rho_\text{out}^{ct} = & \frac{1}{d^4} \sum_{i,j} \ketbra{\Phi_{ij}}^{ct} \notag \\
= & \frac{\id^c}{2} \otimes \frac{\id^t}{d} + \frac{1}{2} \big[ \oprod{0}{1}^c + \oprod{1}{0}^c \big] \otimes T \rho_\text{in}^t T^\dagger \label{eq:rho_out}
\end{align}
where $T\coloneqq\frac{1}{d^2}\sum_iU_i$ and $\rho_\text{in}^t\coloneqq\ketbra{\psi_\text{in}}^t$.
By linearity, Eq.~\eqref{eq:rho_out} also holds for arbitrary mixed inputs $\rho^t_\text{in}$, and the setup thus gives rise to the global channel $\M$ mapping $\rho^t_\text{in}\to\rho^{ct}_\text{out}$.

It is immediately clear that $\rho^{ct}_\text{out}$ depends in general on $\rho_\text{in}^t$, and thus some information can be transmitted through the setup.
If, on the other hand, one classically controls which channel is applied to the input, no information can be transmitted.
Indeed, if the initial state of the control qubit is diagonal, or if it decoheres, in the Pauli $\sigma_z$ eigenbasis, then one can easily check that all dependence on $\rho_\text{in}^t$ disappears in $\rho^{ct}_\text{out}$.
Thus, the global channel $\M$ arising from coherently controlling between $\N_0$ and $\N_1$ provides a communication advantage over classical control.

This advantage mirrors that found using the quantum switch in Ref.~\cite{ebler18}, where the order in which $\N_0$ and $\N_1$ are applied is coherently controlled, thereby applying them in an indefinite causal order.
In the example above, however, indefinite causal order plays no role; 
indeed the two channels are used in parallel (in the two arms of the interferometer), rather than one after the other. 
Nevertheless, the fact that coherent control---of channel or of order---plays a role in transmitting information through fully depolarising channels in both these scenarios raises questions about the role of indefinite causal order in the result of Ref.~\cite{ebler18} using the quantum switch, or, at least, the relative role of causal indefiniteness and coherent control~\cite{loizeau20}.
Similar questions, as well as the need to clarify how to compare different such advantages, have also been raised recently based on different considerations in Ref.~\cite{guerin19}, and subsequently discussed further in Ref.~\cite{kristjansson19}.

In Ref.~\cite{ebler18}, the authors quantified precisely how much classical information can be transmitted by a single use of the depolarising quantum switch (i.e., its Holevo information~\cite{schumacher97,holevo98}).
In Appendix~\ref{ap:HolevoInfo} we present a lower bound for the Holevo information of the global channel $\M$ defined above.
We find that significantly more information can be transmitted by this setup than with the full depolarising quantum switch (e.g., $\frac{1}{2}\log_2\frac{5}{4}\simeq 0.16$ bits compared to $-\frac{3}{8}-\frac{5}{8}\log_2\frac{5}{8}\simeq 0.05$ bits for a qubit target system).

It was further noted in Ref.~\cite{ebler18} that if one traces out either the control or target system from the output of the depolarising quantum switch one obtains the completely mixed state, and thus information is transmitted solely in the correlations between the control and target states.
In the present example, while it is still true that if the control is traced out the target system is left in the completely mixed state, if one traces out the target from Eq.~\eqref{eq:rho_out} one obtains $\rho^c_\text{out}=\frac{1}{2}(\id^c+\Tr[T\rho_\text{in}^tT^\dagger]\sigma_x^c)$, which still depends on $\rho_\text{in}^t$.
Nevertheless, the control system itself does not contain (at any stage in the interferometer) all of the information about the input target state that gets transmitted to $\rho_\text{out}^{ct}$ (it only contains $\Tr[T\rho_\text{in}^tT^\dagger]$, while $\rho_\text{out}^{ct}$ contains $T\rho_\text{in}^tT^\dagger$).
Note that in the setup of Fig.~\ref{fig1}, just as in the depolarising quantum switch, the subspace on which each channel acts  is, technically, a nontrivial $d$-dimensional ``sector'' of he $2d$-dimensional joint control-target Hilbert space\footnote{That is, the channel $\mathcal{C}_i$ acts on the sector corresponding to the control system being in the state $\ketbra{i}^c$.}~\cite{chiribella19},
and so the channels cannot be strictly said to act trivially on the control system~\cite{oreshkov19}.
Indeed, in both cases the global channel acts to delocalise some of the information about the target into the correlations between the two systems. 
This is conceptually similar to the effect of quantum phase kickback associated with controlled unitary operations~\cite{nielsen11,bisio16}, and its role in such communication advantages is further explored in Refs.~\cite{chiribella19,guerin19,kristjansson19}.

\section{Dependence on channel implementation}\label{sec:implementation_dependence}

The approach employed above of randomising over unitary channels is not, however, the only way to implement a fully depolarising channel.
Recall that in general, a quantum channel $\C$ is defined as a CPTP map, and can be described in terms of a (non-unique) set of Kraus operators $\{K_i\}_i$ satisfying $\sum_iK_i^\dagger K_i=\id$, such that the output of the channel is given by $\C(\rho)=\sum_iK_i\rho K_i^\dagger$ for every density matrix $\rho$~\cite{Kraus83,nielsen11,wilde13}.
Note in particular that $\C$ may be applied to a subsystem in a subspace of some larger Hilbert space, as is the case both in the quantum switch and the scenario of Fig.~\ref{fig1}.
There, however, if the channels $\C_0$ and $\C_1$ are not unitary---or not described, as previously considered, as a randomisation over unitary channels---it is \emph{a priori} unclear how to determine the global channel mapping $\rho^t_\text{in}\to\rho^{ct}_\text{out}$ from the Kraus operators of $\C_0$ and $\C_1$.

One possible approach to doing so is to ``purify'' the channels via (independent) Stinespring dilations~\cite{stinespring55}. 
Any channel $\C$ with Kraus operators $\{K_i\}_i$ can indeed be extended to a unitary operation by introducing an environment in an initial state $\ket{\varepsilon}^e$ and considering the  operation that acts on the system under consideration (in our case, the target) and the environment as $\ket{\psi_\text{in}}^t\otimes\ket{\varepsilon}^e\to\sum_iK_i\ket{\psi_\text{in}}^t\otimes\ket{i}^e\coloneqq\ket{\Phi_\text{out}}^{te}$, where the ket vectors $\ket{i}^e$ are (normalised) orthogonal states of the environment. 
After tracing out the environment, we recover $\Tr_e\ketbra{\Phi_\text{out}}^{te}=\sum_iK_i\ketbra{\psi_\text{in}}^tK_i^\dagger=\C(\ketbra{\psi_\text{in}}^t)$, as required.

In the setup of Fig.~\ref{fig1} where the channels $\C_0$ and $\C_1$ have Kraus operators $\{K_i\}_i$ and $\{L_j\}_j$, respectively, one may therefore purify the channels by introducing two, initially uncorrelated, environments with initial states $\ket{\varepsilon_0}^{e_0}$ and $\ket{\varepsilon_1}^{e_1}$.
Note that the control qubit must then be seen as controlling the action of the purified unitary extensions of the channels not only on the target system, but also on the corresponding environments.
This is nevertheless sensible in the interferometric picture of Fig.~\ref{fig1} where the channels may be seen as black boxes with ``internal'' environments, that a photon traverses (in a superposition of ``here'' and ``there'').
  
Under these controlled, purified channels, the combined control-target-environments state evolves unitarily as
\begin{align}
& \ket{+}^c \otimes \ket{\psi_\text{in}}^t \otimes \ket{\varepsilon_0}^{e_0} \otimes \ket{\varepsilon_1}^{e_1} \notag \\
& \to \frac{1}{\sqrt{2}} \ket{0}^c \otimes \sum_i K_i \ket{\psi_\text{in}}^t \otimes \ket{i}^{e_0} \otimes \ket{\varepsilon_1}^{e_1} \notag \\[-1mm]
& \qquad + \frac{1}{\sqrt{2}} \ket{1}^c \otimes \sum_j L_j \ket{\psi_\text{in}}^t \otimes \ket{\varepsilon_0}^{e_0} \otimes \ket{j}^{e_1}. \label{eq:psi_out_purif}
\end{align}
After tracing out the environments, the resulting joint control-target state $\rho_\text{out}^{ct}$ is found to be 
\begin{align}
\rho_\text{out}^{ct} = & \frac{1}{2}\big[ \oprod{0}{0}^c\otimes \C_0(\rho_\text{in}^t) + \oprod{1}{1}^c\otimes\C_1(\rho_\text{in}^t) \big]\notag\\
 &+ \frac{1}{2} \big[ \oprod{0}{1}^c \otimes T_0 \rho_\text{in}^t T_1^\dagger + \oprod{1}{0}^c \otimes T_1 \rho_\text{in}^t T_0^\dagger \big] \label{eq:rho_out_v2}
\end{align}
with $T_0\coloneqq\sum_i\iprod{\varepsilon_0}{i}K_i$ and $T_1\coloneqq\sum_j\iprod{\varepsilon_1}{j}L_j$.

The output state~\eqref{eq:rho_out}, obtained when $\C_0$ and $\C_1$ are depolarising channels implemented as a classical randomisation over $d^2$ orthogonal unitary operators $U_i$, is recovered by taking $K_i=\frac{1}{d}U_i$, $L_j=\frac{1}{d}U_j$, and the initial states of the environment to be $\ket{\varepsilon_0}^{e_0}=\sum_{i=0}^{d^2-1}\frac{1}{d}\ket{i}^{e_0}$, $\ket{\varepsilon_1}^{e_1}=\sum_{j=0}^{d^2-1}\frac{1}{d}\ket{j}^{e_1}$.
Note, however, that a different choice of orthogonal unitary operations (even due to the addition of a relative phase between them, so that, taken individually, they would still implement the same local unitary channels) would have led to a different output state in Eq.~\eqref{eq:rho_out_v2}.
If we had instead taken the environments to initially be in the states $\ket{0}^{e_0}$ and $\ket{0}^{e_1}$ and chosen a set of orthogonal unitary operators such that $K_0=L_0=\frac{1}{d}\id$, we would have obtained Eq.~\eqref{eq:rho_out} with $T=\frac{1}{d}\id$---which, incidentally, coincides with the state of Eq.~\eqref{eq:rho_out_full_switch} obtained in Ref.~\cite{ebler18} as the output of the depolarising quantum switch.
We nevertheless emphasise that Eq.~\eqref{eq:rho_out_v2} gives the output control-target state not only when the channels $\C_0$ and $\C_1$ are obtained by classical randomisation over unitary channels, but for any description of the channels in terms of Kraus operators.

The crucial observation here is that $\rho^{ct}_\text{out}$ depends on the implementation of the channels $\C_0$ and $\C_1$~\cite{aberg04,oi03}.
The interferometric circuit in Fig.~\ref{fig1} is therefore not fully defined by the channels $\C_0$ and $\C_1$, or the Kraus operators chosen to represent them.
This may appear surprising given that, in the usual paradigm, quantum channels are understood to be fully characterised by their (non-unique) Kraus representation~\cite{Kraus83,nielsen11,wilde13}.
However, one should note that such a description of a channel is unchanged under addition of any global phase.
On the other hand, any such ``global'' phase applied by one of the channels in Fig.~\ref{fig1} is only applied to the corresponding arm of the interferometer and therefore, in the overall controlled circuit, becomes a ``relative'' phase with physical significance.
In the case where the channels $\C_0$ and $\C_1$ are unitary, the fact that Fig.~\ref{fig1} is only defined up to such a phase on the unitaries is well known~\cite{zhou13,bisio16}.

\begin{figure}[t]
	\begin{center}
	\includegraphics[width=\columnwidth]{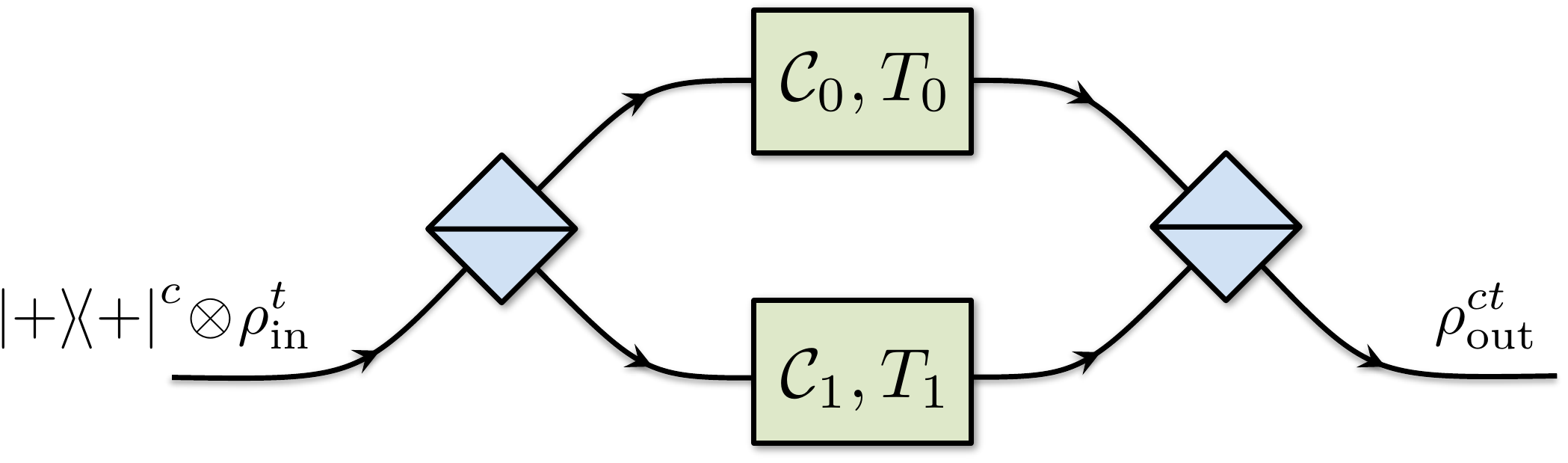}
	\end{center}
	\caption{A corrected version of Fig.~\ref{fig1} in which the description of the two operations inside the interferometer, implementing the channels $\C_0$ and $\C_1$ on their respective subspaces, have been supplemented by the transformation matrices $T_0$ and $T_1$ needed to fully specify the output state $\rho_\text{out}^{ct}$.}
	\label{fig2}
\end{figure}

What we see here, however, is that the output of the interferometric circuit depends not only on any relative phases between (the Kraus operators of) the two channels, but also on a more detailed description of the implementation of the channels. 
More precisely, one requires some additional information encoded in the matrices $T_0$, $T_1$ introduced in Eq.~\eqref{eq:rho_out_v2} in order to fully specify the global channel $\M[\C_0,T_0,\C_1,T_1]:\rho^t_\text{in}\to\rho^{ct}_\text{out}$ induced by the circuit; see Fig.~\ref{fig2}. We call these the ``transformation matrices'' of the channel implementations.
In the description above in terms of a Stinespring dilation, these depend not only on the set of Kraus operators used to decompose the channel, but also on how these are combined (with coefficients that depend on the environment states) to define
$T_0\coloneqq\sum_i\iprod{\varepsilon_0}{i}K_i$ and $T_1\coloneqq\sum_j\iprod{\varepsilon_1}{j}L_j$.\footnote{Note that while $T_0$ and $T_1$ are both independently defined with respect to a common phase reference (supposed to be shared by the two ``boxes'' containing the channels $\C_0, \C_1$ and their environments), only their relative phase is in fact needed to obtain the output state $\rho_\text{out}^{ct}$.}
Let us emphasise that as any channel can be purified in a form that is equivalent to a Stinespring dilation~\cite{Kraus83,nielsen11,wilde13}, the description above is fully general. 

In Appendix~\ref{ap:characterisation} (see Eq.~\eqref{eq:TC}), we characterise completely the transformation matrices $T$ obtainable from some realisation of any given channel $\mathcal{C}$, by deriving a general constraint expressed in terms of the Choi representations~\cite{choi75} of $\C$ and $T$.
For a $d$-dimensional fully depolarising channel, for instance, this constraint simplifies to $\Tr[T^\dagger T]\le\frac{1}{d}$. Under this constraint, applied to both $T_0$ and $T_1$, Eq.~\eqref{eq:rho_out_v2} characterises all possible output states that one can obtain from the setup of Fig.~\ref{fig2}, for any implementation of the channels $\C_0,\C_1=\N$.\footnote{Note that this implies that for certain implementations of depolarising channels the global channel $\M[\N_0,T_0,\N_1,T_1]$ cannot transmit any information; indeed, this is the case if $T_0=0$ or $T_1=0$. 
}

Finally, recalling that the quantum switch is a ``quantum supermap''~\cite{chiribella08a} we note that the transformation $\mathcal{S}[\C_0,\C_1]$ it induces has no such dependence on the implementation of $\C_0$ and $\C_1$, something which is more generally true in any setup in which each channel is applied once and only once to the target system irrespective of the state of the control (see Appendix~\ref{ap:fs}).
The coherent control of quantum channels is thus not a quantum supermap.
We note, however, that Ref.~\cite{chiribella19} has recently introduced the notion of ``vacuum-extended channels'' as an alternative approach to capturing the implementation-dependence relevant to the coherent control of quantum channels.
There, coherent control can be expressed as a supermap of the \emph{vacuum-extended} channels corresponding to a particular implementation, allowing one to formally prove the absence of indefinite causal order (i.e., in that case, the causal separability) of the coherent control of such channels.

\section{Distinguishing different implementations of coherently-controlled channels}

The dependence of the output of the circuit of Fig.~\ref{fig2} on the implementation of the channels means that it is also possible to differentiate between two distinct implementations of the same quantum channel with different transformation matrices.

To see this, consider the case where the channel $\C_0$ has a single, fixed implementation with a transformation matrix $T_0$, while the channel $\C_1$ can have two different possible implementations, with $T_1\neq T_1'$. 
The global channels $\M_{T_1}\coloneqq\M[\C_0,T_0,\C_1,T_1]$ and $\M_{T_1'}\coloneqq\M[\C_0,T_0,\C_1,T_1']$ thus differ in general.
If $T_1$ and $T_1'$ are equally probable, then the maximal probability of successfully distinguishing the two channels---and thereby the two implementations of $\C_1$---is $\frac{1}{2}(1+\mathcal{D}(\M_{T_1},\M_{T_1'}))$, where $\mathcal{D}(\M_{T_1},\M_{T_1'})\coloneqq \frac{1}{2}\lVert \M_{T_1}-\M_{T_1'} \rVert_\diamond$ is the diamond-norm distance between the two global channels~\cite{watrous18}.
In Appendix~\ref{ap:distinguishability} we show that
\begin{align}
\mathcal{D}(\M_{T_1},\M_{T_1'}) \le \frac{1}{2} \big\lVert T_1 - T_1' \big\rVert_2
\label{eq:trDistEqs}
\end{align}
(where $\lVert\cdot\rVert_2$ is the spectral norm),
and that this upper bound can be reached with $\C_0={\cal I}$, $T_0=\id$, in which case it is obtained by taking the input state $\rho_\text{in}^t=\ketbra{\psi_\text{in}}$ maximising $\bra{\psi_\text{in}}(T_1-T_1')^\dagger(T_1-T_1')\ket{\psi_\text{in}}$.
One then discriminates the channels by performing optimal state discrimination between the corresponding output states $\rho_\text{out}^{ct}$ and $\rho_\text{out}^{ct\,\prime}$ of the two global channels.

It is indeed well-known, for instance, that the interferometric setup of Fig.~\ref{fig2} allows one to perfectly discriminate whether the lower arm applies the operation $\ket{\psi}^t\to\ket{\psi}^t$ or $\ket{\psi}^t\to-\ket{\psi}^t$ (the unitaries $\pm \id$), even though these both correspond to the identity channel $\C_1={\cal I}$ (but with $T_1,T_1'=\pm\id$) on the relevant subspace.
As another, perhaps more interesting, example consider the case where $\C_1=\N$ is the fully depolarising channel, with the two possible transformation matrices $T_1^{(\prime)}=\pm\frac{1}{\sqrt{d}}\ketbra{0}$. We have $\frac{1}{2}\big\lVert T_1-T_1'\big\rVert_2=\frac{1}{\sqrt{d}}$, so that these two implementations of the depolarising channel can be distinguished (for $\C_0 ={\cal I}$, $T_0=\id$, $\rho_\text{in}^t=\ketbra{0}^t$) with probability $\frac{1}{2}(1+\frac{1}{\sqrt{d}})$ ($\simeq 0.85$ for $d=2$).

\section{Discussion}\label{sec:discussion}

Coherent control of quantum channels was previously shown to be a resource for communicating through noisy channels in the technique of ``error filtration''~\cite{gisin05}.
Our analysis, following that of Ref.~\cite{ebler18}, shows how coherent control of channels provides more general communication advantages, increasing the capacity of transmission in the absence of postselection and even in the extreme case of completely depolarising channels.
This information gain could be useful in more general error correction and mitigation scenarios, both for quantum communication and computation.

While we focused on depolarising channels to illustrate the ability for two coherently controlled zero-capacity channels to transmit information, this is not the only case where one should intuitively expect no communication to be possible.
Indeed, any constant channel has zero capacity~\cite{schumacher97,holevo98} and similar conclusions can be drawn for any such channel.
Furthermore, while this situation allows for the communication of \emph{classical} information, we note that the authors of Ref.~\cite{ebler18} also recently investigated the transmission of \emph{quantum} information through a quantum switch that puts two dephasing channels in a superposition of orders~\cite{salek18}.
In fact the advantage found there with the quantum switch is also present in the analogous scenario of Fig.~\ref{fig2}; see Appendix~\ref{ap:CoherentInfo}.
Nevertheless, a related recent result has also shown that, for certain choices of noisy channels, the quantum switch can enable \emph{perfect} quantum communication, but that this impossible with our setup~\cite{chiribella18}.
Understanding when the quantum switch or the coherent control of channels is a more powerful communication resource~\cite{loizeau20}, as well as the role of indefinite causal order in communication advantages provided by the quantum switch~\cite{kristjansson19}, are crucial questions that remain the be fully answered.

Our analysis illuminated the fact that the global transformation implemented by the circuit in Fig.~\ref{fig2} depends on the implementation of whatever channels are used.
This stands in contrast to the usual paradigm of quantum channels, where a channel is defined as a CPTP map, and where all descriptions in terms of Kraus operators, or all purifications of a quantum channel, are equivalent~\cite{Kraus83,nielsen11,wilde13}.
Although such a description suffices if a channel is only ever used in isolation, by exploiting quantum control (something possible when the channel is supplied as a ``black-box'' or a usable communication channel~\cite{Araujo14a}) it is in fact possible to extract information about how a channel is implemented, opening up the possibility to use coherent control as a tool for, e.g., error correction~\cite{devitt2013quantum}, quantum channel security~\cite{ambainis2000} and characterisation~\cite{oi03}.

Our results thus show that the notions of coherently controlling quantum channels---and, by extension, their actions when composed in circuits---is, by itself, ill-defined.
Nevertheless, the setups in Figs.~\ref{fig1}--\ref{fig2} that we have considered in this work are perfectly realisable experimentally; indeed, they are in fact less demanding than the approaches already used to implement the full quantum switch~\cite{procopio15,rubino17,rubino17a,goswami18,goswami18a,wei19,guo20}.
Our observations here add to the call (e.g., in Ref.~\cite{Araujo14a} for the control of unknown unitaries) for a generalisation of the standard paradigm of quantum circuits to describe experimentally conceivable situations, that would include the possibility for operations to be quantum-controlled (a general quantum ``if statement''), or more generally to be applied on subspaces only. 
In the situation we considered, we saw that (generalised) quantum channels could be defined not only by the CPTP maps they induce, but also required one to specify the ``transformation matrices'' $T$ introduced above.
We expect that this approach can be used (possibly as a complement to that recently proposed in Ref.~\cite{chiribella19}) for more general situations than the one investigated here~\cite{dong19}, and leave its possible generalisation for future work.

\paragraph{Acknowledgements}

	We thank {\v C}aslav Brukner, Giulio Chiribella, Philippe Allard Gu{\'e}rin, and Hl\'{e}r Kristj\'{a}nsson  for comments on this manuscript.
	We acknowledge financial support from the Swiss National Science Foundation (NCCR SwissMAP and Starting Grant DIAQ) and the \emph{``Investissements d'avenir''} (ANR-15-IDEX-02) program of the French National Research Agency.

\nocite{apsrev41Control} 
\bibliographystyle{apsrev4-2_tweaked}
\bibliography{bib_controlled_channels}

\vspace{1cm}

\onecolumngrid
\appendix

\renewcommand{\theequation}{A\arabic{equation}}
\setcounter{equation}{0}

% \section*{Appendix}

\section{Characterisation of the possible transformation matrices $T$ of a channel $\C$}
\label{ap:characterisation}

Here we characterise completely the transformation matrices $T$ that can be obtained by some implementation of a given channel $\C$, before presenting some examples for specific channels of interest.
To this end, we first recall some details about the Choi isomorphism which will allow us to concisely state and prove our characterisation.

\subsection{Choi isomorphism}

For any given operator $T : \HS_I \to \HS_O$ from some input Hilbert space $\HS_I$ to some output Hilbert space $\HS_O$ (which, for simplicity, we both take to be finite-dimensional), one can define its \emph{Choi vector} representation~\cite{choi75} as
\begin{align}
\dket{T} \coloneqq \id \otimes T \dket{\id} = \sum_m \ket{m} \otimes T\ket{m} \ \in \HS_I \otimes \HS_O,
\end{align}
where $\{\ket{m}\}_m$ is a fixed orthonormal basis of $\HS_I$ and $\dket{\id} \coloneqq \sum_m \ket{m} \otimes \ket{m}$.
Reciprocally, it is easy to see that given a Choi vector $\dket{T} \in \HS_I \otimes \HS_O$, one can recover its corresponding operator $T : \HS_I \to \HS_O$ as
\begin{align}
T = \sum_{m,n} \langle m,n \dket{T} \oprod{n}{m} , \label{eq:Choi_vector_inverse}
\end{align}
where $\{\ket{n}\}_n$ is now a fixed orthonormal basis of $\HS_O$, and where $\bra{m,n} = \bra{m} \otimes \bra{n}$.

In a similar way, one can define the \emph{Choi matrix} representation of any given channel $\C$ from $\L(\HS_I)$ to $\L(\HS_O)$ (with $\L(\HS)$ denoting the space of linear operators over the Hilbert space $\HS$) as
\begin{align}
& C \coloneqq {\cal I} \otimes \C (\dketbra{\id}) = \sum_{m,m'} \oprod{m}{m'} \otimes \C(\oprod{m}{m'})  \in \L(\HS_I \otimes \HS_O),
\end{align}
where ${\cal I}$ is the identity channel.
The channel $\C$ is recovered from its Choi matrix as follows:
\begin{align}
\C (\rho) = \Tr_{\HS_I} [C \cdot (\rho^\text{T} \otimes \id^{\HS_O})] \quad \forall \,\rho,
\end{align}
where $\Tr_\HS$ generically denotes the partial trace over a Hilbert space $\HS$, $\text{T}$ denotes transposition in the chosen basis $\{\ket{m}\}_m$, and where for clarity the superscript in $\id^{\HS_O}$ indicates the Hilbert space on which the identity operator acts.
The fact that $\C$ is, by definition, a completely positive map ensures that its Choi matrix $C$ is positive semidefinite; the fact that $\C$ is trace-preserving implies that $\Tr_{\HS_O} C = \id^{\HS_I}$.

One can easily see that if the channel $\C$ is characterised by a set of Kraus operators $\{K_i\}_i$ (i.e., such that $\C(\rho) = \sum_i K_i \rho K_i^\dagger \ \forall \rho$), then their Choi representations satisfy
\begin{align}
C = \sum_i \dketbra{K_i}. \label{eq:decomp_C_Choi}
\end{align}
Written in this form, it is indeed clear that $C$ is Hermitian positive semidefinite; the constraint that $\Tr_{\HS_O} C = \id^{\HS_I}$ is equivalent to $\sum_i K_i^\dagger K_i = \id$.
We further note that all the $\dket{K_i}$'s are necessarily in the range of $C$: $\dket{K_i} \in \range(C) \ \forall \,i$.%
\footnote{This can easily be seen, e.g., by introducing the projector $\Pi_\C^\perp$ onto the orthogonal complement of $C$, and by noting that $\Tr(\Pi_\C^\perp C) = 0 = \sum_i \Tr(\Pi_\C^\perp \dketbra{K_i})$ implies (as each $\Tr(\Pi_\C^\perp \dketbra{K_i}) \ge 0$) that $\Pi_\C^\perp \dket{K_i} = 0$ for all $i$.}

\subsection{General constraints on the transformation matrices}

Consider a channel $\C$, with its Choi matrix $C$, and let us denote by $C^+$ the Moore-Penrose pseudoinverse of $C$~\cite{barata12}.
As $C$ is positive semidefinite, its pseudoinverse is also positive semidefinite and can be obtained as follows: diagonalising $C$ in the form%
\footnote{Note that this diagonalisation of $C$ is of the form of Eq.~\eqref{eq:decomp_C_Choi}, so that the operators $C_k = \sum_{m,n} \langle m,n \dket{C_k} \oprod{n}{m}$ thus obtained define valid (``canonical'') Kraus operators for the channel $\C$.}
$C = \sum_k \dketbra{C_k}$, with nonzero orthogonal eigenvectors $\dket{C_k}$, one simply has $C^+ = \sum_k \frac{\dketbra{C_k}}{\dbraket{C_k}^2}$.

In the case where $T = \sum_i \iprod{\varepsilon_0}{i} K_i$ is the transformation matrix for a given implementation of the channel $\C$, we have that its Choi vector $\dket{T} = \sum_i \iprod{\varepsilon_0}{i} \dket{K_i} \in \range(C)$ satisfies
\begin{align}
\dbra{T} C^+ \dket{T}^2 & = \Big| \sum_i \iprod{\varepsilon_0}{i} \dbra{T} C^+ \dket{K_i} \Big|^2 \notag \\
& \leq \Big( \sum_i \big| \iprod{\varepsilon_0}{i} \big|^2 \Big) \Big( \sum_i \big| \dbra{T} C^+ \dket{K_i} \big|^2 \Big) \notag \\
& \leq \sum_i \dbra{T} C^+ \dket{K_i} \! \dbra{K_i} C^+ \dket{T} \notag \\
& \qquad = \dbra{T} C^+ C C^+ \dket{T} = \dbra{T} C^+ \dket{T}, \label{eq:proof_charact_T}
\end{align}
where in the second line we used the Cauchy-Schwartz inequality, in the third line we used the fact (due to the normalisation of $\ket{\varepsilon_0}$) that $\sum_i |\iprod{\varepsilon_0}i|^2 \le 1$, and in the fourth line we made use of Eq.~\eqref{eq:decomp_C_Choi} and of the fact that the Moore-Penrose pseudoinverse satisfies $C^+ C C^+ = C^+$.
From Eq.~\eqref{eq:proof_charact_T} it then follows that
\begin{align}
\dbra{T} C^+ \dket{T} \le 1.
\end{align}

Conversely, suppose that an operator $T$ satisfies $\dket{T} \in \range(C)$ and $\dbra{T} C^+ \dket{T} \le 1$. We will see that such a $T$ is the transformation matrix obtained from a particular implementation of the channel $\C$.
Consider indeed the diagonalisation $C = \sum_k \dketbra{C_k}$ introduced already, and define the coefficients $\varepsilon_k \coloneqq \frac{\diprod{C_k}{T}}{\dbraket{C_k}}$. By assumption one has $\sum_k |\varepsilon_k|^2 = \dbra{T} \sum_k \frac{\dketbra{C_k}}{\dbraket{C_k}^2} \dket{T} = \dbra{T} C^+ \dket{T} \le 1$, so that the $\varepsilon_k$'s define valid (subnormalised) coefficients, allowing us to define the initial state $\ket{\varepsilon_0}$ of the environment such that $\iprod{\varepsilon_0}{k} = \varepsilon_k$. One then finds
\begin{align}
\sum_k \iprod{\varepsilon_0}{k} \dket{C_k} & = \sum_k \frac{\dketbra{C_k}}{\dbraket{C_k}} \dket{T} = C C^+ \dket{T} = \dket{T}
\end{align}
(where we used the assumption that $\dket{T} \in \range(C)$ and the fact that $C C^+$ is the orthogonal projector onto $\range(C)$). Equivalently, $\sum_k \iprod{\varepsilon_0}{k} C_k = T$, so we see that the transformation matrix for the particular implementation of the channel $\C$ obtained from the Kraus operators $\{C_i\}_i$ and the initial state $\ket{\varepsilon_0}$ of the environment specified above, and using the Stinespring dilation technique as in the main text is indeed $T$, as desired.

From these observations, one can thus characterise the set ${\cal T}_\C$ of all possible transformation matrices $T$ of a given channel $\C$ as
\begin{align}
{\cal T}_\C = \big\{ T : \dket{T} \in \range(C) \ \text{and} \ \dbra{T} C^+ \dket{T} \le 1 \big\}. \label{eq:TC}
\end{align}

\subsection{Examples}

The Choi matrix of an identity channel ${\cal I}$ is $I = \dketbra{\id}$ (in any dimension); its range is the span of $\dket{\id}$ only, and its Moore-Penrose pseudoinverse is $I^+ = \frac{\dketbra{\id}}{\dbraket{\id}^2} = \frac{\dketbra{\id}}{d^2}$. Eqs.~\eqref{eq:TC} and~\eqref{eq:Choi_vector_inverse} imply that
\begin{align}
{\cal T}_{\cal I} = \big\{ T = \alpha \id \, : \,\alpha \in \mathbb{C}, |\alpha| \le 1 \big\}. \label{eq:Ts_I}
\end{align}
Any such $T = \alpha \id$ with $|\alpha| \le 1$ can indeed be obtained by taking for instance $\{K_i\}_i = \{K_0 = \id\}$ and $\iprod{\varepsilon_0}{0} = \alpha$.
As one can see, even the identity channel does not define a unique transformation matrix. The freedom one has on its possible transformation matrices is not just due to a possible global phase (which would just restrict $\alpha$ above to $|\alpha| = 1$), but also to the possible coherent control of some operation $\ket{\psi}^t \otimes \ket{\varepsilon}^e \to \ket{\psi}^t \otimes \ket{0}^e$ that (while acting trivially on the target system) acts nontrivially on the environment.
Note that Eq.~\eqref{eq:Ts_I} generalises straightforwardly to any unitary channel ${\cal U} : \rho \to U \rho U^\dagger$, whose possible transformation matrices are of the form $T = \alpha U$ with $|\alpha| \le 1$.

\medskip

For a $d$-dimensional fully depolarising channel $\N$, the Choi matrix is $N = \frac{1}{d} \id$; its range is the full Hilbert space $\HS_I \otimes \HS_O$, and its Moore-Penrose pseudoinverse is $N^+ = d \, \id$. Noting that $\dbraket{T} = \Tr[T^\dagger T]$, Eq.~\eqref{eq:TC} implies that
\begin{align}
{\cal T}_\N = \Big\{ T \, : \, \Tr[T^\dagger T] \le \frac{1}{d} \Big\}. \label{eq:Ts_N}
\end{align}
Any such $T$ satisfying $\Tr[T^\dagger T] \le \frac{1}{d}$ can indeed be obtained by taking for instance the set of Kraus operators $\{\frac{1}{d} U_i\}_{i=0}^{d^2-1}$ (where the $U_i$'s are again orthogonal unitary matrices) and $\ket{\varepsilon}^e$ such that $\iprod{\varepsilon_0}{i} = \diprod{U_i}{T} = \Tr[U_i^\dagger T]$.

\medskip

Combining the channels ${\cal I}$ and $\N$, the Choi matrix of a partially depolarising channel (as considered, e.g., in Ref.~\cite{ebler18}) $\N_{(q)} \coloneqq q \, {\cal I} + (1-q) \, \N$, with $0 \le q \le 1$, is $N_{(q)} = q \, I + (1-q) \, N = q \, \dketbra{\id} + (1-q) \, \frac{1}{d} \id$. For $q < 1$ its range is again the full Hilbert space $\HS_I \otimes \HS_O$; its Moore-Penrose pseudoinverse then coincides with its inverse, and is found to be $N_{(q)}^+ = \frac{d}{1-q} [ \id - \frac{d q}{d^2 q +1-q} \dketbra{\id}]$.
Noting that $\diprod{\id}{T} = \Tr[T]$, Eq.~\eqref{eq:TC} then gives
\begin{align}
{\cal T}_{\N_{(q)}} = \Big\{ T \, : \, \Tr[T^\dagger T] - \frac{d q}{d^2 q +1-q} \big|\Tr[T]\big|^2 \le \frac{1-q}{d} \Big\}.
\end{align}
Any $T$ satisfying the above constraint can for instance be obtained with the Kraus operators $\{K_0 = \frac{\sqrt{d^2 q+1-q}}{d} \id, K_i = \frac{\sqrt{1-q}}{d} U_i \text{ for } 1 \le i \le d^2-1\}$ (using a set of orthogonal unitaries that contains $U_0 = \id$) and $\iprod{\varepsilon_0}{0} = \frac{1}{\sqrt{d^2 q+1-q}} \Tr[T]$, $\iprod{\varepsilon_0}{i} = \frac{1}{\sqrt{1-q}} \Tr[U_i^\dagger T]$ for $1 \le i \le d^2-1$.

\medskip

Consider, as another example, the partially dephasing qubit channel in the Pauli $\sigma_z$ eigenbasis---or phase-flip channel---$\Z_{(p)}(\rho) \coloneqq (1-p) \, \rho + p \, \sigma_z \rho \sigma_z$ with $0 \le p \le 1$ (which is completely dephasing for $p = \frac{1}{2}$). Its Choi matrix representation is $Z_{(p)} = (1-p) \, \dketbra{\id} + p \, \dketbra{\sigma_z}$; for $0 < p < 1$ its range is the span of $\{\dket{\id}, \dket{\sigma_z}\}$ and its Moore-Penrose pseudoinverse is $Z_{(p)}^+ = \frac{1}{4(1-p)} \, \dketbra{\id} + \frac{1}{4p} \, \dketbra{\sigma_z}$. Eq.~\eqref{eq:TC} leads to the characterisation (also valid for $p = 0$ or $1$)
\begin{align}
{\cal T}_{{\Z}_{(p)}} & = \big\{ T = \alpha \sqrt{1-p} \,\id + \beta \sqrt{p} \,\sigma_z : \alpha, \beta \in \mathbb{C}, |\alpha|^2 + |\beta|^2 \le 1 \big\}. \label{eq:Ts_Zp}
\end{align}
Any such $T = \alpha \sqrt{1-p} \,\id + \beta \sqrt{p} \,\sigma_z$ with $|\alpha|^2 + |\beta|^2 \le 1$ can straightforwardly be obtained, for instance, from the Kraus operators $\{K_0 = \sqrt{1-p}\,\id, K_1 = \sqrt{p}\,\sigma_z\}$, by taking $\iprod{\varepsilon_0}{0} = \alpha$ and $\iprod{\varepsilon_0}{1} = \beta$.

Similarly, the set of possible transformation matrices for the partially dephasing qubit channel in the $\sigma_x$ eigenbasis---or bit-flip channel---$\X_{(p)}(\rho) \coloneqq (1-p) \, \rho + p \, \sigma_x \rho \sigma_x$ is
\begin{align}
{\cal T}_{\X_{(p)}} & = \big\{ T = \alpha \sqrt{1-p} \,\id + \beta \sqrt{p} \,\sigma_x : \alpha, \beta \in \mathbb{C}, |\alpha|^2 + |\beta|^2 \le 1 \big\}. \label{eq:Ts_Xp}
\end{align}
and any such $T = \alpha \sqrt{1-p} \,\id + \beta \sqrt{p} \,\sigma_x$ can be obtained from the Kraus operators $\{K_0 = \sqrt{1-p}\,\id, K_1 = \sqrt{p}\,\sigma_x\}$, with $\iprod{\varepsilon_0}{0} = \alpha$ and $\iprod{\varepsilon_0}{1} = \beta$.

\section{Communication of classical and quantum information through coherently controlled depolarising and dephasing channels}
\label{ap:capacities}

\renewcommand{\theequation}{B\arabic{equation}}
\setcounter{equation}{0}

\subsection{Holevo information of the coherently controlled depolarising channels}
\label{ap:HolevoInfo}

In this section we look at how much (classical) information can be transmitted by coherently controlling two depolarising channels (i.e., in the scenario of Fig.~\ref{fig2} with $\C_i=\N_i$).
Recall that the Holevo information of a channel $\C$ quantifies how much classical information can be transmitted through a single use of $\C$ from a party $A$ to another party $B$.
It is defined as $\chi(\C) \coloneqq \max_{\{p_a, \rho_a\}} I(A;B)_\nu$, where $I(A;B)_\nu$ is the quantum mutual information calculated on the state $\nu \coloneqq \sum_a p_a \ketbra{a}_A \otimes \C(\rho_a)_B$~\cite{schumacher97,holevo98} (i.e., $I(A;B)_\nu= H(A)_\nu + H(B)_\nu - H(AB)_\nu$, where $H(X)_\nu$ is the von Neumann entropy of the system $X\in\{A,B,AB\}$ in the state $\nu$).
$\chi(\C)$ provides a lower bound for the classical capacity of a quantum channel $\C$.

For a fixed initial state of the control qubit $\rho^c_\text{in} = \ketbra{+}^c$, the depolarising quantum switch can be seen as a global channel, which we denote here $\mathcal{S}[\N_0,\N_1]$, mapping $\rho^t_\text{in}$ to $\rho_\text{out}^{ct}$ (see inset of Fig.~\ref{fig1} in main text), with $\rho_\text{out}^{ct} = \mathcal{S}[\N_0,\N_1](\rho^t_\text{in})$ given by Eq.~\eqref{eq:rho_out_full_switch} in the main text.%
\footnote{Note that as we consider a fixed initial state of the control qubit, which the sender cannot control, it is not part of the input of the channel $\rho_\text{in}^t \to \rho^{ct}_\text{out}$ considered here (but it contributes to the definition of the channel).}
In Ref.~\cite{ebler18} it was shown that the Holevo information of the depolarising quantum switch (with $\rho^c_\text{in} = \ketbra{+}^c$) is $\chi(\mathcal{S}[\N_0,\N_1]) = -\frac{3}{8} -\frac{5}{8} \log_2 \frac{5}{8} \simeq 0.05$ for a qubit target system, while a more general formula for any dimension $d$ was also given.

The situation of Fig.~\ref{fig2} similarly induces a global channel, which we denote by $\M[\C_0,T_0,\C_1,T_1]$, mapping $\rho^t_\text{in}$ to $\rho_\text{out}^{ct}$ according to Eq.~\eqref{eq:rho_out_v2} of the main text.
A lower bound on the Holevo information $\chi(\M[\C_0,T_0,\C_1,T_1])$ can (for a given pair of channels and transformation matrices) be easily obtained by simply taking $I(A;B)_\nu$, for any particular choice of the weighted ensemble $\{p_a,\rho_a\}$.
For two fully depolarising channels $\N_0, \N_1$ with $T_0 = T_1 =\frac{1}{\sqrt{d}}\ketbra{0}$ for instance (which indeed satisfies $\Tr[T_i^\dagger T_i] \le \frac{1}{d}$ as required by Eq.~\eqref{eq:Ts_N}), taking $\{p_0 = \frac{3}{5}, \rho_0 = \ketbra{0}, p_1 = \frac{2}{5}, \rho_1 = \ketbra{1}\}$ gives the lower bound $\chi(\M[\N_0,T_0,\N_1,T_1]) \ge \frac{1}{d}\log_2\frac{5}{4}$, which is a significant increase over that obtained by the depolarising quantum switch. For $d = 2$ this indeed gives $\chi(\M[\N_0,T_0,\N_1,T_1]) \gtrsim 0.16$; for larger $d$ this bound decreases as the dimension increases, but less rapidly than the (exact) Holevo information obtained in Ref.~\cite{ebler18}.
It remains an open question whether our lower bound is tight, both for the transformation matrices $T_i$ as well as for any other transformation matrices for two completely depolarising channels.

One may be tempted to argue that the fact that the lower bound obtained here exceed the Holevo information for the depolarising quantum switch obtained in Ref.~\cite{ebler18} is perhaps not so surprising given the differences between the scenario in Fig.~\ref{fig2} and that of the quantum switch.
Indeed, in the scenario we consider, the target system only goes through the depolarising channels (in a superposition) a single time, while in the quantum switch the target system always goes through both channels (in a superposition of different orders).
Thus, one may intuitively expect the target system to be less ``degraded'' in the scenario considered here.
However, a recent paper exploring numerically this issue~\cite{loizeau20} (which appeared subsequent to an earlier preprint of this manuscript and, in part, in response to the very questions we raise) indicates that the situation is more subtle, with the noncommutativity of the Kraus operators of the two channels placed inside a quantum switch also playing a role in these communication advantages.
In particular, in some cases the quantum switch indeed outperforms coherent control meaning this intuition---at least applied naively---is incorrect, although further study is needed to understand fully the situation.

\subsection{Quantum information transfer through coherently controlled dephasing channels}
\label{ap:CoherentInfo}

Here we look at how much quantum information can be transmitted through two coherently controlled complementary dephasing channels.
Recall that the quantum information that can be communicated through a channel $\C$ from some system $A$ to some other system $B$ is quantified by the quantum capacity $\Q(\C)$~\cite{lloyd97,shor02,devetak05}. 
A lower bound on $\Q(\C)$ is given by the coherent information $\Q^{(1)}(\C) \coloneqq \max_{\nu_0} [H(B)_\nu - H(AB)_\nu]$, where the maximisation is over all states $\nu_0 \in \L(\HS_A \otimes \HS_{A'})$ of a bipartite system comprising $A$ and a reference system $A'$ isomorphic to $A$, and with $\nu \coloneqq (\mathcal{I} \otimes \C)(\nu_0) \in \L(\HS_A \otimes \HS_B)$~\cite{schumacher97}. 
A further lower bound on $\mathcal{Q}^{(1)}(\C)$ can of course be obtained by choosing any specific state $\nu_0$.

Consider, as in Ref.~\cite{salek18}, the case where the channels $\C_0$ and $\C_1$ in Fig.~\ref{fig2} are (two-dimensional) phase-flip and the bit-flip channels $\mathcal{Z}_{(p)}$ and $\mathcal{X}_{(p)}$, respectively (for simplicity we take the same mixing parameter $p$ for both channels).
The possible transformation matrices for these channels are characterised by Eqs.~\eqref{eq:Ts_Zp} and~\eqref{eq:Ts_Xp}. Taking $T_{\Z_{(p)}} = \sqrt{p} \, \sigma_z$ and $T_{\X_{(p)}} = \sqrt{p} \, \sigma_x$ (i.e., taking $\alpha = 0, \beta = 1$ in Eqs.~\eqref{eq:Ts_Zp}--\eqref{eq:Ts_Xp}), and writing $\M_{(\Z,\X,p)} \coloneqq \M[\Z_{(p)},T_{\Z_{(p)}},\X_{(p)},T_{\X_{(p)}}]$, we find, from Eq.~\eqref{eq:rho_out_v2} of the main text, that 
\begin{align}
&\M_{(\Z,\X,p)}(\rho_\text{in}^t)  =  (1-p) \frac{\id}{2} \otimes \rho_\text{in}^t + p \,\frac{\ket{0}^c \otimes \sigma_z + \ket{1}^c \otimes \sigma_x}{\sqrt{2}} \rho_\text{in}^t \frac{\bra{0}^c \otimes \sigma_z + \bra{1}^c \otimes \sigma_x}{\sqrt{2}}.
\end{align}

This allows one to calculate $\nu \coloneqq ({\cal I} \otimes \M_{(\Z,\X,p)})(\nu_0)$, and then $H(B)_\nu - H(AB)_\nu$, for any choice of $\nu_0$.
By taking the maximally entangled state $\nu_0 = \ketbra{\Phi^+}$ with $\ket{\Phi^+} \coloneqq \frac{1}{\sqrt{2}}(\ket{0}\otimes\ket{0} + \ket{1}\otimes\ket{1})$, after some calculation we obtain
\begin{align}
\Q(\M_{(\Z,\X,p)})) & \ge \Q^{(1)}(\M_{(\Z,\X,p)}))  \ge p - H_2(p) + H_2\big(\frac{1-p}{2}\big),
\end{align}
where $H_2$ is the binary entropy function $H_2(p)\coloneqq -p\, \log_2 (p) - (1-p)\log_2(1-p)$.

This lower bound on the quantum capacity is larger than that obtained in Ref.~\cite{salek18} for the full quantum switch with phase-flip and bit-flip channels.
Remarkably, our bound is positive for all values of $p$---in particular, it takes the value $-\frac{3}{4}\log_2\frac{3}{4} \simeq 0.31$ for $p = \frac{1}{2}$, i.e., for fully dephasing channels (which, by themselves, cannot transmit any quantum information).
It is also larger than the quantum capacities of each channel $\Z_{(p)}$ and $\X_{(p)}$ individually---and thus violates the ``bottleneck inequality'' considered in Ref.~\cite{salek18}---for all $p$.
In comparison, the bound obtained in Ref.~\cite{salek18} with the full quantum switch was positive only for $p \lesssim 0.13$ and $p \gtrsim 0.60$, and was larger than $\Q(\Z_{(p)})$ or $\Q(\X_{(p)})$ only for $p \gtrsim 0.62$.

As for the Holevo information, we remark that, in our scenario, the target system only ever passes through a single dephasing channel, which may, at least in part, help explain the higher bound we obtain on the quantum capacity.
Once again, however, the situation merits further study, as shown by the results of Ref.~\cite{chiribella18} mentioned in Sec.~\ref{sec:discussion} which show that the quantum switch can, in some cases, enable perfect quantum communication where coherent control cannot.

\section{Channel implementation independence for the full quantum switch and other quantum processes}
\label{ap:fs}

\renewcommand{\theequation}{C\arabic{equation}}
\setcounter{equation}{0}

For the case of the full quantum switch, each of the two channels $\C_0$ and $\C_1$, with Kraus operators $\{K_i\}_i$ and $\{L_j\}_j$, is applied once and only once on the target system. 
Considering a purified version of the channels via a Stinespring dilation, as described in the main text, the state at the output of the interferometer (see the inset of Fig.~\ref{fig1}) reads
\begin{align}
& \frac{1}{\sqrt{2}} \ket{0}^c \otimes \sum_{i,j} L_j K_i \ket{\psi_\text{in}}^t \otimes \ket{i}^{e_0} \otimes \ket{j}^{e_1}  + \frac{1}{\sqrt{2}} \ket{1}^c \otimes \sum_{i,j} K_i L_j \ket{\psi_\text{in}}^t \otimes \ket{i}^{e_0} \otimes \ket{j}^{e_1}.
\end{align}
In contrast to the output state~(5) for the circuit of Fig.~\ref{fig1}, no terms appear in which either environment is untouched and remains in its initial state. After tracing out the environments, one obtains
\begin{align}
\rho_\text{out}^{ct} = & \frac{1}{2} \Big( \ketbra{0}^c \otimes \,\C_1\circ\C_0(\rho_\text{in}^t) +  \ketbra{1}^c \otimes \C_0\circ\C_1(\rho_\text{in}^t)  +  \oprod{0}{1}^c \otimes \sum_{i,j} L_j K_i \rho_\text{in}^t L_j^\dagger K_i^\dagger  + \oprod{1}{0}^c \otimes \sum_{i,j} K_i L_j \rho_\text{in}^t K_i^\dagger L_j^\dagger \Big),
\end{align}
which depends neither on the initial state of the environments, nor on the sets of Kraus operators chosen to describe each channel. 
Indeed, for any other Kraus representations $\{M_r\}_r$ of $\C_0$ and $\{N_s\}_s$ of $\C_1$, one has $K_i = \sum_r u_{ir} M_r$ and $L_j = 
\sum_s v_{js} N_s$, where $u_{ir}$ and $v_{js}$ are the elements of unitary matrices~\cite{nielsen11}. We thus obtain 
\begin{align}
 \sum_{i,j} L_j K_i \rho_\text{in}^t L_j^\dagger K_i^\dagger & = \sum_{i,j} \sum_{r,r',s,s'} u_{ir}u_{ir'}^* v_{js} v_{js'}^* N_s M_r \rho_\text{in}^t N_{s'}^\dagger M_{r'}^
\dagger \notag\\
& = \sum_{r,r',s,s'} \delta_{r,r'} \delta_{s,s'} N_s M_r \rho_\text{in}^t N_{s'}^\dagger M_{r'}^\dagger =  \sum_{r,s} N_s M_r \rho_\text{in}^t N_{s}^\dagger M_{r}^\dagger
\end{align}
(where $\delta$ is the Kronecker delta),
and analogously for the term $\sum_{i,j} K_i L_j \rho_\text{in}^t K_i^\dagger L_j^\dagger$.

\medskip

More generally, consider a combination of multiple channels $\C_0, \ldots, \C_{N}$ with Kraus operators $\{K^{(0)}_{i_0}\}_{i_0}, \ldots, \{K^{(N)}_{i_{N}}\}_{i_{N}}$, and assume that for any possible initial state $\ket{\Psi_\text{in}}$ sent through the setup, each channel is applied once and only once (not necessarily in a definite order).
Considering a Stinespring dilation of the channels with environment initial states $\ket{\varepsilon_0}^{e_0}, \ldots, \ket{\varepsilon_{N}}^{e_{N}}$, this means that the joint initial state evolves as
\begin{align}
& \ket{\Psi_\text{in}} \otimes \ket{\varepsilon_0}^{e_0} \otimes \cdots \otimes \ket{\varepsilon_{N}}^{e_{N}}  \to  \sum_{i_0, \ldots, i_{N}}  F(K^{(0)}_{i_0},\ldots, K^{(N)}_{i_{N}}) \ket{\Psi_\text{in}} \otimes  \ket{i_0}^{e_0} \otimes \cdots \otimes \ket{i_{N}}^{e_{N}}, \label{eq:transfo_N_channels}
\end{align}
where each $F(K^{(0)}_{i_0},\ldots, K^{(N)}_{i_{N}})$ is an operator composed as a sum of product terms in which each $K^{(\ell)}_{i_\ell}$ appears once and only once, in possibly different orders (e.g.,\ for the quantum switch: $\ket{\Psi_\text{in}} = \ket{+}^c \otimes \ket{\psi_\text{in}}^t$ and $F(K_i,L_j) = \ketbra{0}^c \otimes L_j K_i + \ketbra{1}^c \otimes K_i L_j$).
For any such transformation, a similar calculation as for the full quantum switch can be conducted, which shows that after tracing out the environments, the final output state does not depend on the choice of Kraus operators, nor on the initial states of the environments.

We note that the assumption that each channel is applied---or that ``each party acts''---once and only once is at the core of the process matrix framework~\cite{oreshkov12,wechs18a}. This justifies, beyond the particular case of the quantum switch, that the situations described by process matrices do not depend on any specific implementation of the channels (or of more general quantum operations) applied by each party, but only on the description (e.g.,\ in terms of Kraus operators) of the induced completely positive maps.

\section{Distinguishing different implementations of coherently-controlled channels}
\label{ap:distinguishability}

\renewcommand{\theequation}{D\arabic{equation}}
\setcounter{equation}{0}

In this section we show that the diamond-norm distance in the scenario of Fig.~\ref{fig1}, where the channel $\C_0$ has a fixed transformation matrix $T_0$ while the channel $\C_1$ has two possible transformation matrices $T_1$ and $T_1'$, is obtained and bounded as in Eq.~\eqref{eq:trDistEqs} of the main text.

Recall that the diamond-norm distance between the two channels $\M_{T_1}\coloneqq\M[\C_0,T_0,\C_1,T_1]$ and $\M_{T_1'}\coloneqq\M[\C_0,T_0,\C_1,T_1']$ is defined as 
\begin{align}
	\mathcal{D}(\M_{T_1},\M_{T_1'}) &\coloneqq \frac{1}{2}\big\lVert \M_{T_1}-\M_{T_1'} \big\rVert_\diamond = \frac{1}{2} \max_{\rho^{tr}_\text{in}} \big\lVert (\M_{T_1}\otimes\mathcal{I})(\rho^{tr}_\text{in})-(\M_{T_1'}\otimes\mathcal{I})(\rho^{tr}_\text{in}) \big\rVert_1,
\end{align}
where $\lVert\cdot\rVert_1$ is the trace norm and the maximum is taken over all joint states $\rho^{tr}_\text{in}$ of the target system ($t$) and some arbitrary reference system ($r$).
That is, it is the maximum trace-distance between output states when the channels are applied to part of joint system (and with the identity channel $\mathcal{I}$ being applied to the other part); indeed, considering an input state entangled with a reference system is in general necessary for optimal channel discrimination~\cite{watrous18}.

Eq.~\eqref{eq:rho_out_v2} can straightforwardly be adapted to give the corresponding output state $\rho^{ctr}_\text{out}$ of the control, target and reference systems, by replacing $\rho^{t}_\text{in}$ by $\rho^{tr}_\text{in}$, $\C_{0/1}$ by $\tilde \C_{0/1} \coloneqq \C_{0/1}\otimes\mathcal{I}$, and $T_{0/1}$ by $\tilde T_{0/1} \coloneqq T_{0/1} \otimes \id$. Defining $\tau_1 \coloneqq T_1 - T_1'$ and $\tilde \tau_1 \coloneqq \tau_1 \otimes \id = \tilde T_1 - \tilde T_1'$ for ease of notation, we have, for an arbitrary input state $\rho^{tr}_\text{in}$,
 \begin{align}
\frac{1}{2}\big\lVert (\M_{T_1}\otimes\mathcal{I})(\rho^{tr}_\text{in})-(\M_{T_1'}\otimes\mathcal{I})(\rho^{tr}_\text{in}) \big\rVert_1 & = \frac{1}{2} \Tr \sqrt{(\rho_\text{out}^{ctr} - \rho_\text{out}^{ctr\,\prime})^2} \notag \\
 & = \frac{1}{4} \Tr \sqrt{
 \ketbra{0}^c \otimes \tilde T_0 \rho_\text{in}^{tr} \tilde \tau_1^\dagger \tilde \tau_1\rho_\text{in}^{tr} \tilde T_0^\dagger + \ketbra{1}^c \otimes \tilde \tau_1\rho_\text{in}^{tr} \tilde T_0^\dagger \tilde T_0 \rho_\text{in}^{tr} \tilde \tau_1^\dagger
 } \notag \\
 & = \frac{1}{4} \Big( \big\lVert \tilde \tau_1\rho_\text{in}^{tr} \tilde T_0^\dagger \big\rVert_1 + \big\lVert \tilde T_0 \rho_\text{in}^{tr} \tilde \tau_1^\dagger \big\rVert_1 \Big) = \frac{1}{2} \, \big\lVert \tilde \tau_1\rho_\text{in}^{tr} \tilde T_0^\dagger \big\rVert_1, \label{eq:normSimplified}
 \end{align}
where the last equality follows from the invariance of the trace norm under Hermitian transposition.

Decomposing $\rho^{tr}_\text{in}$ in the form $\rho^{tr}_\text{in} = \sum_m p_m \ketbra{\psi_m}$ (with $p_m \ge 0$, $\sum_m p_m = 1$, $\braket{\psi_m} = 1 \ \forall m$), using the sub-additivity of the trace norm, and the fact that $\lVert \oprod{x}{y} \rVert_1 = \sqrt{\braket{x}} \sqrt{\braket{y}}$ for any two vectors $\ket{x}$, $\ket{y}$, we further have
\begin{align}
\frac{1}{2}\,\big\lVert \tilde \tau_1\rho^{tr}_\text{in} \tilde T_0^\dagger \big\rVert_1 & \le \frac{1}{2} \sum_m p_m \, \big\lVert \tilde \tau_1 \ketbra{\psi_m} \tilde T_0^\dagger \big\rVert_1 = \frac{1}{2} \sum_m p_m \, \sqrt{\bra{\psi_m}\tilde \tau_1^\dagger\tilde\tau_1\ket{\psi_m}}  \sqrt{\bra{\psi_m}\tilde T_0^\dagger \tilde T_0\ket{\psi_m}}.
\end{align}
Now, by the definition of the spectral norm one has $\sqrt{\bra{\psi_m}\tilde \tau_1^\dagger\tilde\tau_1\ket{\psi_m}} \le \lVert \tilde\tau_1 \rVert_2 = \lVert \tau_1 \rVert_2$ for all $m$. Additionally,
\begin{align}
	& \bra{\psi_m} \tilde T_0^\dagger \tilde T_0 \ket{\psi_m}  =  \Big(\sum_{i'}  \big(\bra{\psi_m}K_{i'}^\dagger \otimes \id^r \big) \otimes \bra{i'}^{e_0}\Big)\Big(\id^{t}\otimes\id^{r} \otimes \ketbra{\epsilon_0}^{e_0}\Big)\Big(\sum_{i} \big(K_{i} \otimes \id^r \ket{\psi_m} \big) \otimes \ket{i}^{e_0}\Big)\le 1.
\end{align}
Combined together, these bounds lead directly to the upper bound of Eq.~\eqref{eq:trDistEqs}.
Furthermore, it is easy to see that the inequalities above---and thus the upper bound of Eq.~\eqref{eq:trDistEqs}---can always be saturated by taking $T_0 = \id$ (with $\C_0 = {\cal I}$) and $\rho^{tr}_\text{in}=\rho_\text{in}^t = \ketbra{\psi}$ with $\ket{\psi}$ maximising $\bra{\psi}\tau_1^\dagger\tau_1\ket{\psi}$, i.e., without any (or with a trivial) reference system.

\medskip

To finish with, we note that in the example presented in the main text in which $\C_1 = \N$, the two implementations we proposed, with $T_1, T_1' = \pm \frac{1}{\sqrt{d}} \ketbra{0}$, are the most distinguishable ones that can be considered. Indeed, according to Eq.~\eqref{eq:Ts_N} any transformation matrix $T_1$ of $\C_1 = \N$ must satisfy $\Tr[T_1^\dagger T_1] = \big\lVert T_1 \big\rVert_\text{HS}^2 \le \frac{1}{d}$ (with $\lVert\cdot\rVert_\text{HS}$ denoting the Hilbert-Schmidt norm; this constraint is indeed satisfied for our choice above). Using the fact that the Hilbert-Schmidt norm always upper-bounds the spectral norm, it follows that
\begin{align}
\frac{1}{2} \big\lVert T_1 - T_1' \big\rVert_2 & \le \frac{1}{2} \big\lVert T_1 - T_1' \big\rVert_\text{HS}  \le \frac{1}{2} \big( \big\lVert T_1 \big\rVert_\text{HS} + \big\lVert T_1' \big\rVert_\text{HS} \big)  \le \frac{1}{\sqrt{d}},
\end{align}
so that the value one obtains with the choice above, $\frac{1}{2} \lVert T_1 - T_1' \rVert_2 = \frac{1}{\sqrt{d}}$, is indeed the largest possible one.

\end{document}